\documentclass[showpacs,floatfix,aps,prl,preprint,superscriptaddress]{revtex4-1}
\usepackage{float}
\usepackage{graphicx}
\usepackage{amsmath}
\usepackage[T1]{fontenc}
\usepackage{epstopdf}
\usepackage[caption=false]{subfig}
\usepackage{xfrac}
\usepackage{amssymb}
\usepackage{braket}
\usepackage{graphicx}
\usepackage{verbatim}
\usepackage{etoolbox}
\usepackage{amsfonts} 
\usepackage{amsmath} 
\usepackage[utf8]{inputenc} 
\usepackage{mathtools}
\usepackage{soul,xcolor,colortbl}
\usepackage{lipsum}
\usepackage{hyperref} \hypersetup{colorlinks=true,citecolor=blue,linkcolor=blue,urlcolor=blue} 

\usepackage{ulem}
\usepackage{bm} 
\usepackage{array}
\usepackage{color}

\begin{document}


\title{Quantifying and Visualizing the Microscopic Degrees of Freedom of Grain Boundaries in the Wigner-Seitz Cell of the Displacement-Shift-Complete Lattice}

\author{I. S. Winter}
\email{iswinte@sandia.gov}
\affiliation{Sandia National Laboratories, Albuquerque, NM 87123, USA}
\author{T. Frolov}
\email{frolov2@llnl.gov}
\affiliation{Lawrence Livermore National Laboratory, Livermore, CA 94550, USA}






\date{\today}

\begin{abstract}

We introduce a grain boundary (GB) translation vector, $\bm{t}^{WS}$, to describe and quantify the domain of the microscopic degrees of freedom of GBs. It has long been recognized that for fixed macroscopic degrees of freedom of a GB there exists a large multiplicity of states characterized by different relative grain translations. More recently another degree of freedom, $[n]$, the number of GB atoms, has emerged and is now recognized as an equally important component of GB structural multiplicity. In this work, we show that all GB microstates can be uniquely characterized by their value of $\bm{t}^{WS}$, which is located within the Wigner-Seitz (WS) cell of the Displacement-Shift-Complete lattice (DSCL) of the GB. The GB translation vector captures information about both the translation state and the number of GB atoms.  We show that the density of GB microstates inside the cell of the DSCL is not uniform and can form clusters that correspond to different GB phases. The vectors connecting the centers of the clusters correspond to the Burgers vectors of GB phase junctions, which can be predicted without building the junctions. Using $\bm{t}^{WS}$, we quantify GB excess shear and argue that it is defined up to a DSCL vector, which has implications for thermodynamic equilibrium conditions. Additionally, this work generalizes the definition of the number of GB atoms [n] to asymmetric boundaries.


\end{abstract}

\maketitle

\section{Introduction}

Grain boundaries (GBs), interfaces formed between misoriented crystals of the same material, have five macroscopic degrees of freedom that define the orientation relationship between the crystals and the orientation of the boundary plane \cite{sutton1995interfaces,HOMER2022118006}. By direct analogy with three-dimensional bulk systems, GBs can have microstates and multiple GB phases \cite{Zhu2018}. It is important to distinguish between GB microstates and GB phases. A microstate is a particular configuration of a system at the microscopic level, defined by the positions and momenta of all the particles in the system. A phase, on the other hand, encompasses a large number of similar microstates and has well-defined ensemble-averaged thermodynamic properties, which result in an equation of state. For GB phases, these thermodynamic averages are the excess properties introduced by Gibbs \cite{Gibbs}.

Microscopic degrees of freedom describe the possible variations of the GB structure for fixed macroscopic degrees of freedom. In elemental systems, relative grain translations parallel to the boundary plane, $\bm{t} = t_1 \hat{\bm{e}}_1+t_2\hat{\bm{e}}_2$, and the fractional number of GB atoms, $[n]$, are currently used to navigate this space, as different values of $\bm{t}$ and $[n]$ must result in different GB microstates. For decades, the relative grain translation vector, $\bm{t}$, alone was considered sufficient to sample the space of all relevant GB microstates. The work by Pond quantified this space of distinct allowed translations parallel to the boundary plane \cite{Pond1977a,Pond1977b}. For sufficiently large translations, the resulting structures repeat, forming identical grain boundaries but on different atomic planes \cite{Pond1977b,bollmann2012crystal}. This periodicity facilitates the formation of disconnections, a GB dislocation with an associated step \cite{HIRTH1994985,POND1994287,HIRTH19964749,HAN2018386}, with the Burgers vector residing within the Displacement-Shift-Complete lattice (DSCL) associated with the GB \cite{bollmann2012crystal,GRIMMER19741221}. Unfortunately, grain boundary structure searches restricted to sampling translations parallel to the GB plane alone led to unphysical predictions.

For example, it was claimed that a commonly observed $\langle 100 \rangle$-twist GB in MgO cannot be stable without segregation of charged impurities \cite{Wolf1983} and should fracture spontaneously due to strong Coulombic repulsion between ions with similar charges sitting on top of each other at the GB. This erroneous prediction arose because the number of ions at the boundary was not properly optimized. By simply deleting one of the two same-charge ions in the boundary core, Tasker and Duffy predicted stable and low-energy structures for these boundaries \cite{doi:10.1080/01418618308243118}. Two decades later, another claim emerged that $\langle 100 \rangle$-twist boundaries in silicon could be amorphous at 0 K \cite{PhysRevLett.77.2965,Keblinski}. A subsequent study by von Alfthan et al. predicted ordered solid structures by again optimizing the number of atoms in the GB core in addition to the grain translations parallel to the boundary plane \cite{PhysRevLett.96.055505}.

Despite these studies, the requirement to optimize the number of atoms or ions at GBs was largely attributed to systems with special bonding characteristics, such as covalent or ionic crystals and was still considered to be less important for metallic systems, with their more uniform and delocalized bonding. As a result, for decades it was widely believed that structures of high-angle grain boundaries in metallic systems predicted by empirical potentials become disordered at temperatures as low as half of the melting point \cite{PhysRevB.27.5576,PhysRevB.85.144104,MISHIN20093771,Zhang2009}. These simulations led to the acceptance of the notion that finite-temperature structures of GBs could resemble a glass. However, new GB phases with ordered solid structures were later predicted in elemental metals when the number of atoms was allowed to vary \cite{Frolov2013}. Moreover, it was discovered that GBs of different types in face-centered cubic (FCC) \cite{Frolov2013,Zhu2018}, body-centered cubic \cite{Frolov_nanoscale,FROLOV2018123} and hexagonal close-packed \cite{chen2024grand} metals commonly have multiple phases, composed of different numbers of atoms, and first-order structural transitions between these different ordered states can be observed when the number of atoms changes. These GB structures that require optimization with respect to the number of atoms have been recently observed with high-resolution transmission electron microscopy \cite{doi:10.1126/science.adq4147,seki2023incommensurate}.

Despite this clear motivation, our current understanding of the microscopic degrees of freedom and the space available for distinct microscopic GB states remains incomplete. The relative grain translations and the number of GB atoms, $[n]$, are fundamentally different properties that have been forced to be considered together as GB microscopic degrees of freedom only because their combination successfully addressed the shortcomings of previous simulations. Additionally, grain boundary structure optimization that involves variable numbers of GB atoms is yet to be performed for asymmetric boundaries. This is because asymmetric GBs can contain a different number of atoms per plane in the two crystals, and the current definition of $[n]$ needs to be generalized to accomodate this fact.

In this study, we show that the grain boundary microstates should be considered inside the Wigner-Seitz (WS) cell of the DSCL of the boundary and that these states can be identified in this space by the GB translation vector, $\bm{t}^{WS}$, the components of which capture both the translations parallel to the boundary plane as well as $[n]$. Quantifying this space allows us to predict how many atoms should be inserted or deleted in any boundary type, including asymmetric boundaries, before the structures repeat. By introducing a generalized GB translation vector and visualizing the density of states of the vector within the WS cell of the DSCL, we can predict line defects formed by different grain boundary phases based on the individual boundary structures without simulating actual transformations and defects. Further, the generalized GB translation vector addresses the practical need for GB structure prediction algorithms to ensure that the entire structure space is explored. We show that this translation vector can also serve as an additional descriptor, in addition to GB excess properties, for the classification of GB phases.

\section{ Defining Grain Boundary Translation Vectors}

\begin{figure}[ht!]
\centering
\includegraphics[width=0.4\columnwidth]{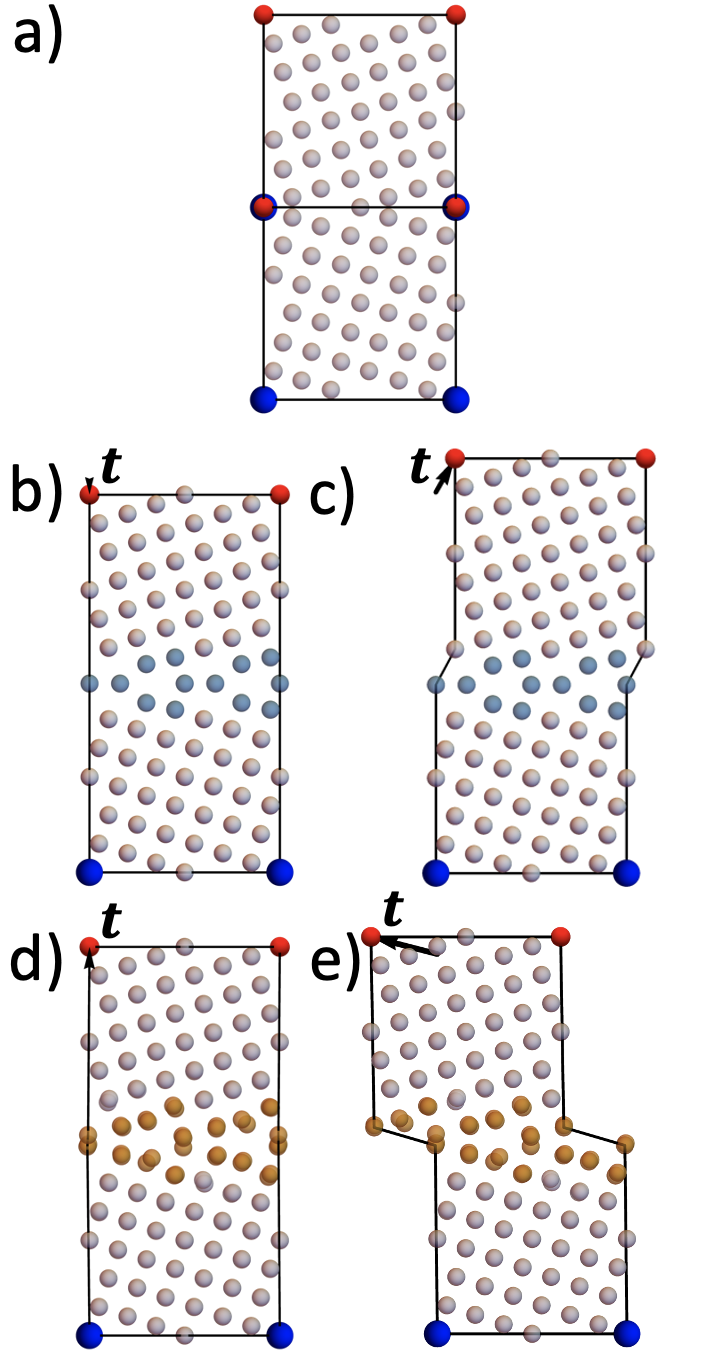}
\caption{An illustration of how two different GB microstates with regular kite and split kite structures are created out of the reference bulk crystals by applying different relative translations and by inserting or removing atoms in the GB core. Panel a: the reference state consists of two misoriented crystals forming a coincidence site lattice without a grain boundary present. Panels b and c: two different bicrystals with the same regular kite structure are created from the reference state (a) by applying different relative grain translations and removing (b) or inserting (c) atoms in the GB core equivalent to an integer number of atomic planes. Panels d and e: two different bicrystals with the same split kite structure are created from the reference state (a) by applying different relative grain translations and  removing (b) or inserting (c) atoms in the GB core not equivalent to an integer number of atomic planes. The arrows indicate the displacement vector $\bm{t}$, describing the change in the relative positions of the bulk lattices from the reference state due to the creation of the GBs.}
\label{fig:kite-schematic}
\end{figure}

We begin the consideration of grain boundary degrees of freedom by considering a process in which different GB structures, or microstates, can be created out of a reference state. The reference state shown in Fig. \ref{fig:kite-schematic} contains two misoriented crystals that form a coincidence site lattice (CSL). In Fig. \ref{fig:kite-schematic}a, the corners of the upper and lower crystals correspond to CSL positions and are shown as  red and blue points. To create a GB from this state we keep the positions of the blue lattice sites fixed while displacing the upper crystal and changing the number of atoms in the GB core. Fig. \ref{fig:kite-schematic}b and \ref{fig:kite-schematic}c show two different bicrystals obtained as a result of this procedure, both containing the same microstate: a regular kite structure. The bicrystals have different shapes and numbers of atoms as shown with the translation vector, $\bm{t}$, that tracks the displacement of the upper crystal relative to the lower crystal. In Fig. \ref{fig:kite-schematic}b, $\bm{t}$ points downward, indicating that a certain number of atoms, equivalent for this particular structure to an integer number of atomic planes, was removed from the GB core during the creation process. By contrast, in Fig. \ref{fig:kite-schematic}c extra atomic planes were inserted. While the regular kite structure can be obtained by simply joining two reference half crystals together, we consider a general process where the GB atoms are also inserted or removed to illustrate that $\bm{t}$ can have different values while resulting in the same atomic structure.

Similarly, Fig. \ref{fig:kite-schematic}d and e show a GB with a microstate, the split kite structure, different from the regular kite structure. Split kite structures, identified in Ref. \cite{Frolov2013} have a different translation state compared to regular kites and also require an insertion (or removal) of $[n]\approx0.5$, with $[n]$ being the fraction of an atomic plane relative to the reference state. In Ref. \cite{Frolov2013} $[n]$ is defined as the ratio of the number of atoms inserted into the GB to create a structure and the number of atoms per atomic plane.  This distinction is important because it shows that split kites cannot be obtained from regular kites by simply rearranging the atoms in the boundary core. The number of atoms at the boundary represents its own degree of freedom, separate from the rigid body translations. It is also evident from Fig. \ref{fig:kite-schematic}d and e that the boundary structure can be preserved by a combination of translations parallel to the boundary and the addition or removal of atomic planes in the GB core, just as in the case of the regular kite structure. 

Any two translation vectors that correspond to the same GB microstate such as those in Figs \ref{fig:kite-schematic}b and c or \ref{fig:kite-schematic}d and e differ from each other by exactly a DSCL vector. Alternatively stated, the lattice formed by all possible translation vectors to create the same microstate corresponds to a DSCL. On the other hand, the translation vectors for different microstates such as regular kites and split kites are not related to each other through a DSCL vector, but have a contribution from the differences in excess GB properties such as excess volume and excess shears. We now derive a relation between the translation vectors, grain boundary excess properties and the DSCL.

Following the analysis from Ref. \cite{FrolovBurgers}, we introduce the deformation gradients $\bm{F}^b$ and $\bm{F}$ that map to the configurations of a bulk grain and the bicrystal respectively from a reference configuration with volume, $V'$, and cross-sectional area, $A'$. The deformation gradients effectively describe the shape of the bulk grain configuration and the bicrystal traced by blue and red markers as shown in Fig. \ref{fig:kite-schematic}. The relation between $\bm{t}$ and the deformation gradients is given by

\begin{equation}\label{eq:tdef}
    t_i^{} = \frac{V'}{A'}(F_{i3}^{}-F^{b}_{i3}),
\end{equation}
with a derivation of Eq. \eqref{eq:tdef} given in the supplemental materials.

As noted in Ref. \cite{PhysRevB.85.224106}, excess shears and volume of a grain boundary are given by the following determinant:

\begin{equation}\label{eq:determinant}
V'[JF_{i3}/F_{33}]_N\; = \frac{V'}{N^b}
\begin{vmatrix}
J F_{i3}/F_{33} & J^b F^b_{i3}/F^b_{33} \\
N & N^b
\end{vmatrix},
\end{equation}
where $J = \det\bm{F}$ and $J^b = \det\bm{F}^b$ are Jacobians, while $N$ and $N^b$ are the number of lattice sites in the bulk and bicrystal configurations. By combining Eqs. \eqref{eq:tdef} and \eqref{eq:determinant} we obtain a relation between the translation vector, GB excess properties and the number of atoms present in the bulk and bicrystal configurations:

\begin{equation}\label{eq:excess-t}
    t_i A = V'[JF_{i3}/F_{33}]_N + \Omega^b(N-N^b) F^b_{i3}/F^b_{33},
\end{equation}
where $\Omega^b$ is the volume per lattice site in the bulk configuration. Eq. \eqref{eq:excess-t} explains how the normal component of the translation vector, $t_3$, is related to the number of atoms that were inserted or removed during the GB creation process as well as the excess volume $[V]_N$ (see the supplemental materials for more detail). As discussed in the supplemental materials, for a cubic system we can define $\bm{F}^b$ such that it is a diagonal matrix. This means that for the cubic system considered in Fig. \ref{fig:kite-schematic}, the excess shear is directly proportional to the vector, $\bm{t}$. As a result, the lateral components of the translation vector are simply equal to the excess shear given by the determinant in Eq. \eqref{eq:determinant}. 

We have previously pointed out that for a given GB microstate the translation vector is defined up to a DSCL vector. The same must be true for the excess shears. We can thus write $t_i = t_i^{WS} + v_i^{DSC}$ and $V'[JF_{i3}/F_{33}]_N = B_i + w_i^{DSC}$, with $\bm{v}^{DSC}$ and $\bm{w}^{DSC}$ being DSCL vectors. Eq. \eqref{eq:excess-t} becomes

 \begin{subequations}\label{eq:tWS-B}
     \begin{align}
         t_1^{WS} &= B_1 + d^{DSC}_1,\\
         t_2^{WS} &= B_2 + d^{DSC}_2,\\
         t_3^{WS} &= [V]_N + n\Omega^b/A + d^{DSC}_3,\label{eq:t3ws}
     \end{align}
 \end{subequations}
 with $\bm{d}^{DSC}$ being a DSCL vector and $n$ refers to the number of extra atoms due to the GB. $\bm{t}^{WS}$ is unique for a given GB structure and characterizes the microscopic degrees of freedom of the structure capturing the translations parallel and normal to the boundary plane. We now illustrate how $\bm{t}^{WS}$ can be calculated for a given GB structure.

\section{Calculating Unique Translation Vectors}

\begin{figure}[ht!]
\centering
\includegraphics[width=\columnwidth]{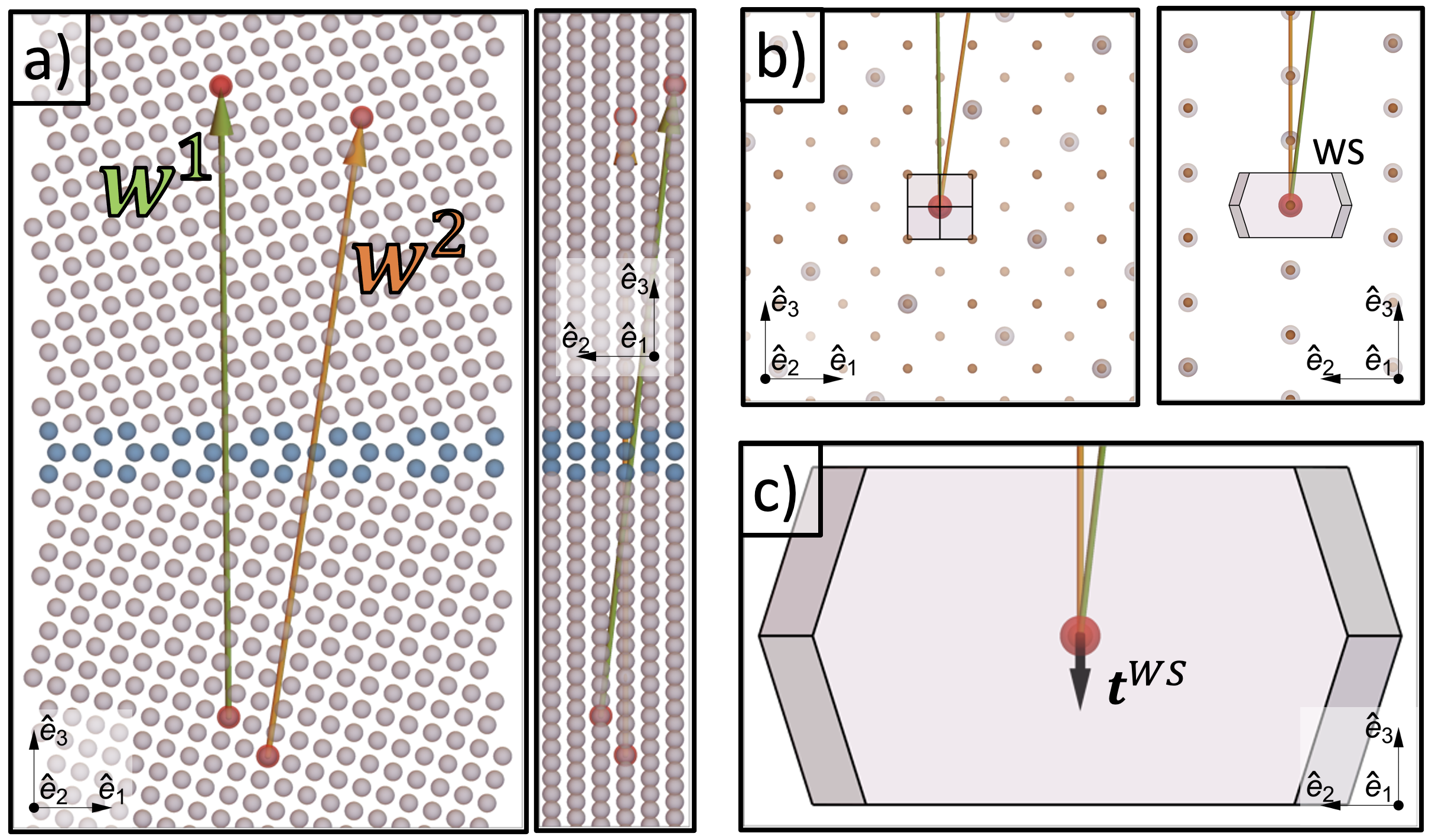}
\caption{An illustration of how a GB translation vector, $\bm{t}^{WS}$, is calculated for a regular kite structure of the $\Sigma 5(310)[001]$ GB in Cu.  Panel a: examples of GB crossing vectors connecting arbitrary lattice sites of the two grains. Panel b: WS cell of the DSCL (brown points) centered on the red lattice site ( $\bm{w}^2$ has been rigidly translated such that it has the same origin as $\bm{w}^1$ in the figure). Panel c: the different crossing vectors are wrapped back into the WS cell to produce the same unique GB translation vector $\bm{t}^{WS}$ shown with a black arrow. }
\label{fig:ws-kite}
\end{figure}

To calculate $\bm{t}^{WS}$, we first draw a GB crossing vector, $\bm{w}$, by connecting two lattice sites on the two sides of the GB, as shown in Fig. \ref{fig:ws-kite} for the case of a $\Sigma 5 (310)[001]$ grain boundary with FCC grains. The lattices of the two grains are referred to as lattice $\mathit{1}$ and $\mathit{2}$. The specific choice of the lattice sites is not important so long as these sites are located sufficiently far enough away from the GB such that they are unaffected by any strain fields originating from the GB. Two possible crossing vectors $\bm{w}^1$ and $\bm{w}^2$ are illustrated in Fig. \ref{fig:ws-kite}. The difference between any two crossing vectors can be expressed as 

\begin{equation}\label{eq:dw-DSC}
     \bm{w}^2 - \bm{w}^1 = \bm{a}^2 - \bm{a}^1 = \bm{d}^{DSC},
\end{equation}
where $\bm{a}^1$ and $\bm{a}^2$ are vectors belonging to lattices $\mathit{1}$ and $\mathit{2}$, respectively. As lattices $\mathit{1}$ and $\mathit{2}$ form a CSL, the right hand side of Eq. \eqref{eq:dw-DSC} is a DSCL vector, $\bm{d}^{DSC}$, by its definition \cite{GRIMMER19741221}. Eq. \eqref{eq:dw-DSC} shows that the set of all possible crossing vectors forms a DSCL. However, because a given $\bm{w}$ crosses the boundary, the crossing vector will consist of a DSCL vector, $\bm{d}$, and a relative translation vector between the two grains, $\bm{t}$, such that we can always write a crossing vector as $\bm{w} = \bm{d} + \bm{t}$.

To specify one unique vector that captures the translational properties of a given GB structure, $\bm{t}^{WS}$, we select the value of $\bm{t}$ that resides within the Wigner-Seitz cell of the DSCL.  The Wigner-Seitz cell of a given lattice point is defined as the region of space closest to that lattice point \cite{ashcroft1976solid}. As such, we can calculate $\bm{t}^{WS}$ by considering a given crossing vector, $\bm{w}$. We generate the DSCL associated with the boundary, such that the DSCL has a lattice point at the origin of $\bm{w}$. We then find the DSCL vector that is the nearest neighbor of $\bm{w}$, denoted as $\bm{d}^{NN}$. We calculate the smallest possible translation vector for the GB structure, the unique translation vector, to be

\begin{equation}\label{eq:define_tau}
    \bm{t}^{WS} = \bm{w}-\bm{d}^{NN}.
\end{equation}



We illustrate the procedure to calculate $\bm{t}^{WS}$ for a $\Sigma 5(310)[001]$ symmetric tilt GB in Cu modeled with an embedded atom method (EAM) potential \cite{PhysRevB.63.224106}. The well known regular kite structure of this boundary is illustrated in Fig. \ref{fig:ws-kite}a. This structure is the lowest energy state found for this particular system when using the $\gamma$-surface approach, which samples  relative  grain translations parallel to the boundary and does not involve insertion or removal of atoms from the boundary plane. Viewing Fig. \ref{fig:ws-kite}a, the kite structure itself appears to contain mirror symmetry along the GB plane. For this reason one may be tempted to assume that $\bm{t}^{WS}$ for this GB structure is composed of the excess volume, $[V]_N$, alone: $\bm{t}^{WS} = [V]_N\hat{\bm{e}}_3$. 


To calculate the unique $\bm{t}^{WS}$ we can consider both crossing vectors: $\bm{w}^1 = -0.572$ \AA $ \hat{\bm{e}}_1 - 5.423$ \AA $  \hat{\bm{e}}_2+ 46.043$ \AA $ \hat{\bm{e}}_3$ and $\bm{w}^2 = 6.859$ \AA $ \hat{\bm{e}}_1 + 46.615$ \AA $ \hat{\bm{e}}_3$ shown graphically in Fig. \ref{fig:ws-kite}. The DSCL for the bicrystal can be expressed in terms of the primitive lattice vectors: $\bm{a}^{DSC} = \frac{a_0}{\sqrt{10}}\hat{\bm{e}}_1$, $\bm{b}^{DSC} = -\frac{a_0}{2\sqrt{10}}\hat{\bm{e}}_1 + \frac{a_0}{2}\hat{\bm{e}}_2+\frac{a_0}{2\sqrt{10}}\hat{\bm{e}}_3$, and $\bm{c}^{DSC} = \frac{a_0}{\sqrt{10}}\hat{\bm{e}}_3$, with $a_0=3.615$ \AA\ for this particular potential.  We note that the primitive lattice vectors of a given DSCL can be calculated using a procedure given in Ref. \cite{ADMAL2022118340}. We then subtract all DSCL contributions from the crossing vectors to find the unique translation vector located within the WS cell of the DSCL. As shown in \ref{fig:ws-kite}c, both crossing vectors collapse into the same vector, $\bm{t}^{WS} = -0.255$ \AA$\hat{\bm{e}}_3$, showing that $\bm{t}^{WS}$ is a unique property of the given GB structure.

The normal component of $\bm{t}^{WS}$ clearly does not match the expected excess volume, $[V]_N = 0.316$\ \AA\ \cite{FrolovBurgers}. To obtain a translation vector more amenable to our intuition, we take $\bm{t}^{WS}$ and add $\bm{b}^{DSC}$ to find $\bm{t} = -\frac{a_0}{2\sqrt{10}} \hat{\bm{e}}_1 + \frac{a_0}{2} \hat{\bm{e}}_2 +[V]_N\hat{\bm{e}}_3$. This translation vector is no longer the shortest, but $t_3$ matches the excess volume. Since the normal component matches the excess volume exactly, the calculated $\bm{t}=\bm{t}^{WS}+\bm{b}^{DSC}$ also predicts that no atoms were inserted or removed from the boundary core during the construction procedure and $[n]=0$. $t_1$ and $t_2$ represent non-zero translations parallel to the boundary plane required to construct the GB structure. When using the $\gamma$-surface approach, we intentionally tracked the relative translations of the grains during the boundary construction process and verified that the relative translation $\bm{t} = -0.572$\ \AA$\hat{\bm{e}}_1 + 1.807$\ \AA$\hat{\bm{e}}_2 + 0.316$\ \AA$\hat{\bm{e}}_3$ is indeed required to produce the regular kite structure.

While the obtained relative translation vector, $\bm{t}^{WS}+\bm{b}^{DSC}$,  exactly matches the relative translation vectors we find in the GB construction procedure, with the Wigner-Seitz-based approach we do not need to know the history of how the boundary was created. As a result, we can extract all possible $\bm{t}$ vectors from the final structure alone, by simply adding DSCL vectors to $\bm{t}^{WS}$. This calculation illustrates how the GB microscopic degrees of freedom, the relative grain translations parallel to the boundary and normal to the boundary due to excess volume and the number of GB atoms, are captured by $\bm{t}^{WS}$.

\section{ Predicting Burgers Vectors for Grain Boundary Phase Junctions}

We now demonstrate that this analysis allows us to predict Burgers vectors of GB phase junctions, without constructing the actual junctions. Consider the $\Sigma 5 (210)[001]$ symmetric tilt GB in Cu modeled with the EAM potential from Ref. \cite{PhysRevB.63.224106}. The potential predicts three distinct phases for this boundary: regular kites, split kites and filled kites \cite{Frolov2013}, the latter two requiring grand-canonical optimization \footnote{Grand canonical optimization of a grain boundary consists of lateral relative translations of the grains as well as insertions and deletions of atoms in the GB} as all three phases have a different number of GB atoms \cite{Zhu2018}. The fractional number of GB atoms, $[n]$, is listed for each phase in Table \ref{tab:tau_kite}. 

First-order structural phase transformations between two phases induced by temperature have been previously demonstrated using MD simulations \cite{Frolov2013}. During the transformation both GB phases are present and are separated by a line defect, referred to as a grain boundary phase junction. The Burgers vector of this junction was quantified in our previous publication using the Burgers circuit construction method \cite{FrolovBurgers}. However, the Burgers circuit construction approach requires the simulation of the actual junction.


\begin{figure}[ht!]
\centering
\includegraphics[width=\columnwidth]{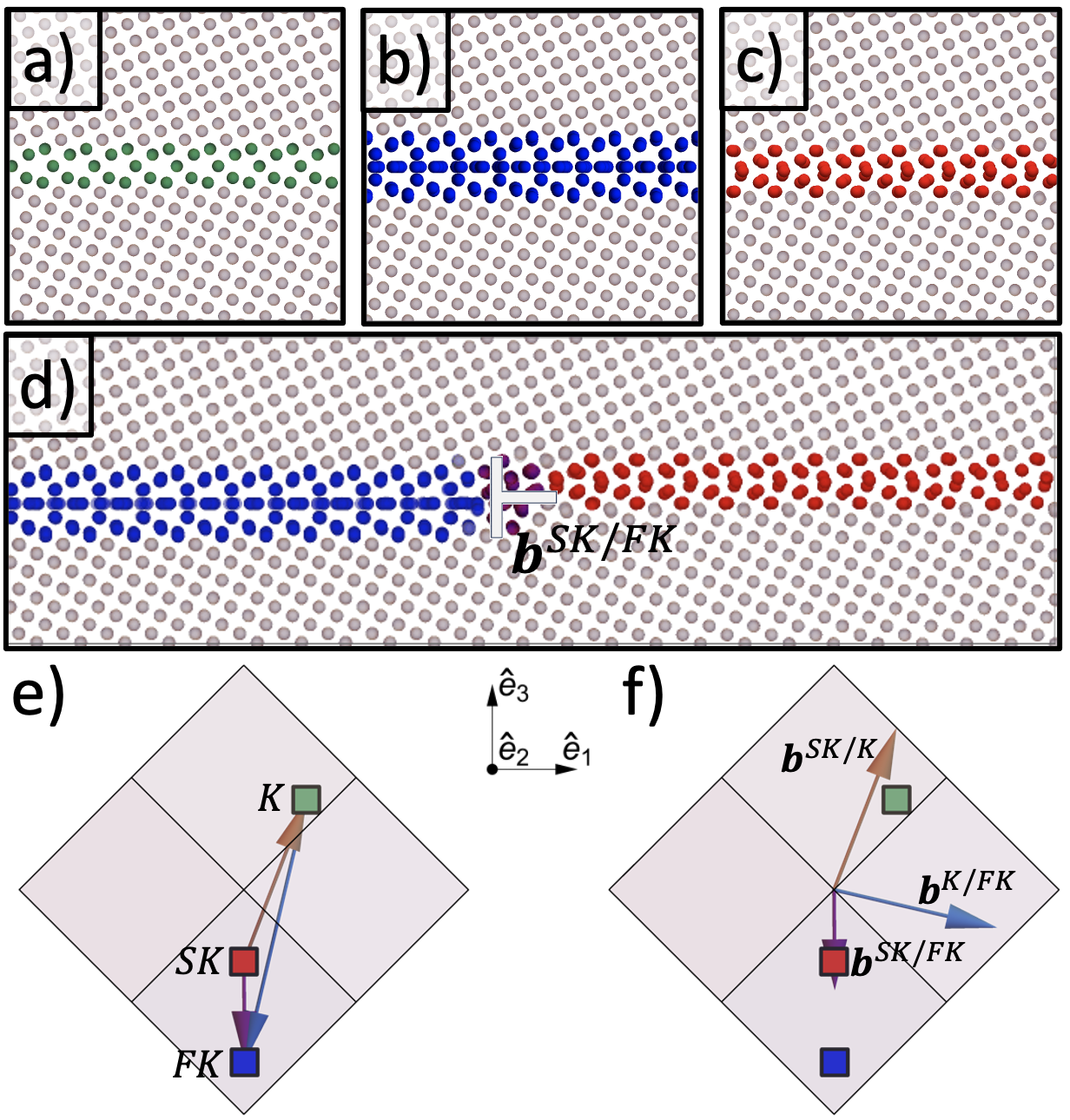}
\caption{Panels a-c: regular kite (K), filled-kite (FK), and split-kite (SK) GB phases of the $\Sigma5(210)[001]$ GB in Cu. Panel d: A GB phase junction between a filled-kite and split-kite phase with the junction colored purple. Panel e: the $\bm{t}^{WS}$ vectors of the three GB phases are shown with squares inside the Wigner-Seitz cell (grey volume) of the displacement-shift-complete lattice. The differences between $\bm{t}^{WS}$ vectors (shown as purple, light blue, and orange vectors connecting the squares) predict possible, but not necessarily the smallest, Burgers vectors of GBPJs between these GB phases. Panel f: the smallest Burgers vector associated with each GB phase junction can be calculated by subtracting DSCL vectors such that the Burgers vectors are wrapped inside the WS cell.}
\label{fig:kite_Burgers}
\end{figure}

\begin{table}[hbt!]
\caption{\label{tab:tau_kite} Excess volume, fractional number of GB atoms and unique translation vectors for three GB phases identified in the $\Sigma 5(210)[001]$ symmetric tilt GB.}
\begin{ruledtabular}\footnotesize
\begin{tabular}{llllll}
Structure & $[V]_N$ (\AA) & $[n]$ & $t^{WS}_1$ (\AA) & $t^{WS}_2$ (\AA) & $t^{WS}_3$ (\AA)\\ \hline
split kite & 0.172 & 0.467 & 0.000  & 0.000 & -0.259 \\
filled kite & 0.301 & 0.857 & 0.003 & 0.001 & -0.621 \\
kite & 0.322 & 0.000 & 0.224 & 0.000 & 0.323
\end{tabular}
\end{ruledtabular}
\end{table}


Table \ref{tab:tau_kite} provides $\bm{t}^{WS}$ for the three different GB phases calculated in this study. These phases are plotted as squares within the WS cell of the DSCL as shown in Fig. \ref{fig:kite_Burgers}. We can use these $\bm{t}^{WS}$ vectors to predict all possible Burgers vectors that can exist at junctions between any two GB phases ($\alpha$ and $\beta$) by taking the difference

\begin{equation}\label{eq:Burgers-vector}
    \bm{b}^{\alpha/\beta} = \bm{t}^{WS,\beta} - \bm{t}^{WS,\alpha} + \bm{d}^{DSC},
\end{equation}
where $\bm{d}^{DSC}$ represents a general DSCL vector. 


In Fig. \ref{fig:kite_Burgers}e we plot the calculated differences in unique translation vectors between the split, filled and regular kite structures as arrows. These arrows are the Burgers vectors of the three possible GB phase junctions. As demonstrated in \ref{fig:kite_Burgers}f, the obtained Burgers vectors $\bm{b}^{SK/FK}=\bm{t}^{WS,FK}-\bm{t}^{WS,SK}$ and $\bm{b}^{SK/K}=\bm{t}^{WS,K}-\bm{t}^{WS,SK}$ reside within the WS cell, and are thus the smallest possible Burgers vectors. However, $\bm{b}^{K/FK}=\bm{t}^{WS,FK}-\bm{t}^{WS,K}$, must be wrapped into the WS cell to find the smallest possible Burgers vector. Figure \ref{fig:kite_Burgers}f shows all three smallest Burgers vectors predicted for these junctions.


We can further illustrate how the microscopic degrees of freedom of individual boundaries captured by $\bm{t}^{WS}$ contribute to the Burgers vector. For example, for the split-kite/filled-kite junction illustrated in Fig. \ref{fig:kite_Burgers}d  we find $\bm{b}^{SK/FK} = 0.003$\ \AA$ \hat{\bm{e}}_1 +0.001$\ \AA$ \hat{\bm{e}}_2  - 0.362$ \AA$\hat{\bm{e}}_3$, using the values given in TABLE \ref{tab:tau_kite}, which matches the value measured using the Burgers circuit approach in our previous study \cite{FrolovBurgers}, but without the need to ever construct or simulate the actual junction.

The tangential components of the Burgers vector correspond to the differences in the relative grain translation parallel to the boundary plane. While the normal component contains contributions from the differences in excess volume and the number of GB atoms. Indeed, by applying Eq. \eqref{eq:tWS-B} the calculated $t_3^{WS}$ components of the two GB phases can be expressed as 
\begin{subequations}
    \begin{align}
        t_3^{WS,SK} &= [V]_N^{SK} + n^{SK} \Omega^b/A + d_3^{DSC},\\
        t_3^{WS,FK} &= [V]_N^{FK} + n^{FK} \Omega^b/A + d_3^{DSC},\\
        b_3^{SK/FK} &= \left([V]_N^{FK}-[V]_N^{SK}\right) + \left(n^{FK}-n^{SK}\right)\Omega^b/A + d_3^{DSC}.
    \end{align}
\end{subequations}
We note that, unlike the regular kite structure of the $\Sigma 5(310)[001]$ GB discussed previously, $t_3^{WS,SK}$ and $t_3^{WS,FK}$ are no longer just excess volumes but contain a non-zero contribution from the number of GB atoms that have to be inserted or deleted within the GB core to create the respective structures. Using the values for excess volume and numbers of atoms from TABLE \ref{tab:tau_kite} we obtain the normal component of the Burgers vector $b_3=0.362$\ \AA$\hat{\bm{e}}_3$. 

\section{GB phases and Distribution of GB microstates inside the WS cell of the DSCL}

Computational grand-canonical structure searches can generate hundreds to thousands of distinct GB structures or microstates of the same boundary. Therefore, there exists a practical need to analyze this vast set of structures and extract meaningful information and identify all distinct GB phases. A previous study proposed using various descriptors that can be calculated for each structure \cite{Zhu2018}. When excess GB properties are used as descriptors, clusters appear in the descriptor space, and individual GB phases can be classified automatically using machine learning techniques \cite{Zhu2018}. Here we show that that the components of $\bm{t}^{WS}$ can be used as such descriptors.

Using the same $\Sigma 5(210)[001]$ GB described earlier, we now consider how the density of the different generated GB microstates  described by $\bm{t}^{WS}$ is distributed inside the WS cell of the DSCL. We show that for this boundary, $\bm{t}^{WS}$ vectors of all generated microstates cluster inside the WS cell, with each cluster of microstates corresponding to one of the three GB phases. 

\begin{figure}[ht!]
\centering
\includegraphics[width=\columnwidth]{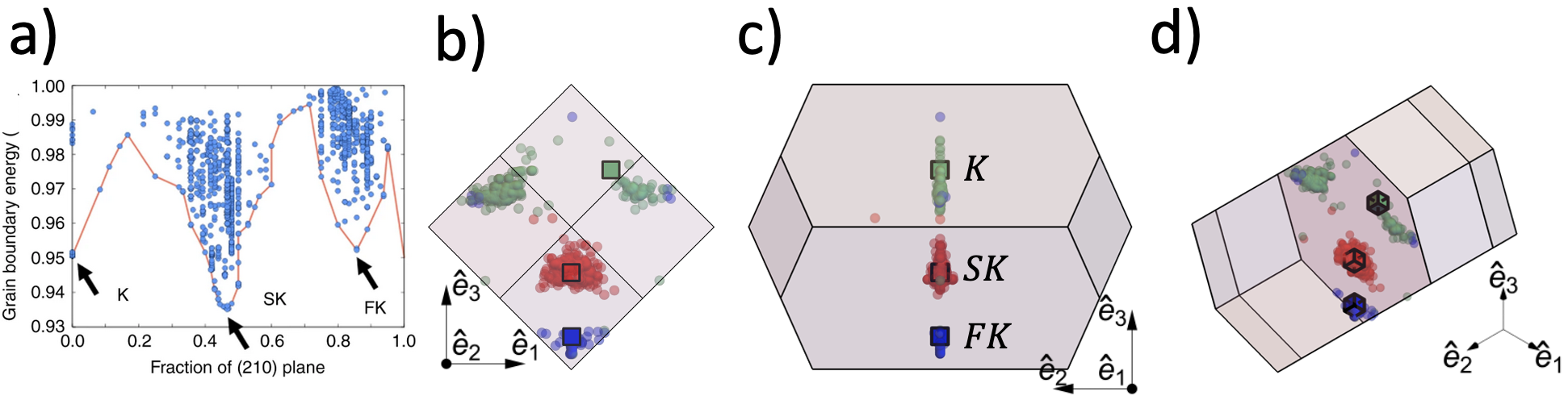}
\caption{Plots of unique translation vectors associated with all structures obtained from a grand-canonical optimization procedure \cite{Zhu2018}. The lowest energy structures for each GB phase are identified with cubes. Panel a: a figure obtained from Ref. \cite{Zhu2018} showing a grand canonical GB structure search at 0 K generates approximately a thousand distinct GB microstates and predicts three distinct lowest energy phases of this boundary: K, SK and FK. Panels b-d: Unique translation vectors, $\bm{t}^{WS}$, are calculated for each structure and plotted inside the WS cell of the DSCL with each panel showing a different viewpoint. The density of microstates is not uniform and shows strong clustering with each of the three clusters corresponding to a different GB phase. The clusters were identified using a K-means algorithm \cite{Zhu2018}.}
\label{fig:cluster_kites}
\end{figure}

We calulate $\bm{t}^{WS}$ for 1013 GB structures generated using the evolutionary structure prediction code in a previous study \cite{Zhu2018}. In that study, the three distinct GB phases were clustered based on properties such as $[V]_N$, GB stress, $[n]$ and Steinhardt order parameters \cite{PhysRevB.28.784,10.1063/1.2977970}. Fig. \ref{fig:cluster_kites} shows the calculated $\bm{t}^{WS}$ of each of the 1013 structures, visualized as points within the WS cell of the DSCL. Each point is colored according to one of the three structure types: regular kite (green), split kite (red), and filled kite (blue), as identified previously in Fig. \ref{fig:kite_Burgers}. Remarkably, we observe that the translation vectors representing the microstates are not distributed uniformly within the WS cell, but form three clusters, each associated with a distinct GB phase. 
As discussed in the previous section, the relative spacings of the three cluster centers are directly related to the Burgers vectors of the GBPJs which  control the GB phase transformation barriers \cite{Winter2022,Winter2024}.

Our approach is of additional practical significance, because $\bm{t}^{WS}$ can potentially be used as a robust, easy to calculate, descriptor for machine learning techniques to identify distinct GB phases and navigate the structure search algorithms in the space of allowed distinct microstates. The robustness of the clustering technique obviously depends on how the values of $\bm{t}^{WS}$ are distributed within the WS cell. If there is no strong clustering, if the Burgers vectors of the possible junctions are too small, it may not be possible to extract unique phases. Moreover, a more uniform distribution of the microstates in the WS cell may suggest critical-point-like behavior of the GB structure when all microstates are easily sampled at high temperature.

\section{Quantifying GB Excess Shear}

Interface excess shear is a thermodynamic excess property that describes the change in the interfacial free energy due to an applied shear stress parallel to the interfacial plane, as described by the adsorption equation for a grain boundary \cite{PhysRevB.85.224106,PhysRevB.85.224107}:

\begin{equation}\label{eq:gibbs}
    d(\gamma A) = -[S]_N dT - A[V]_N d\sigma_{33} - A \sum_{i=1}^2B_{i}d\sigma_{3i} + A\sum_{i,j=1}^2 \tau_{ij} de_{ij},
\end{equation}
where $[S]_N$ is the excess entropy, $\bm{B}$ the excess shear, and $\bm{\tau}$ is the GB stress. In Eq \eqref{eq:gibbs} the two GB excess shears, $B_1$ and $B_2$, are coupled to shear stresses parallel to the boundary plane in the same manner as excess volume, $[V]_N$, is coupled to the normal component of stress. While these three properties are closely linked and describe the response of the boundary to an applied stress, excess volume is routinely calculated and has been reported in numerous studies of interfaces and grain boundaries by both simulations and experiments \cite{BURANOVA2016367,PhysRevLett.108.055504,doi:10.1073/pnas.2400161121,Frolov2013,FrolovBurgers}. Excess shear, on the other hand, remains poorly understood and has been qualified and reported by a handful of studies \cite{PhysRevB.85.224106,PhysRevB.85.224107}. This is despite the fact that one could argue that excess shear has an even greater influence on material behavior than excess volume. The difference in excess shears of two different GB structures allows for dislocations at GBPJs with components along the GB plane. These glissile dislocations can be activated during GB-mediated mechanical deformation \cite{PADMANABHAN2024146713,Wilde2023MT-MF2022009,DILLON2023118718} or GB migration \cite{CAHN20064953,CHEN2019241, GORKAYA20095396,doi:https://doi.org/10.1002/9783527652815.ch09,TRAUTT20126528}.  The major reason why excess shear receives so little attention in the literature is because there is still no clear recipe for how to quantify it for each individual boundary. 

\begin{figure}[ht!]
\centering
\includegraphics[width=\columnwidth]{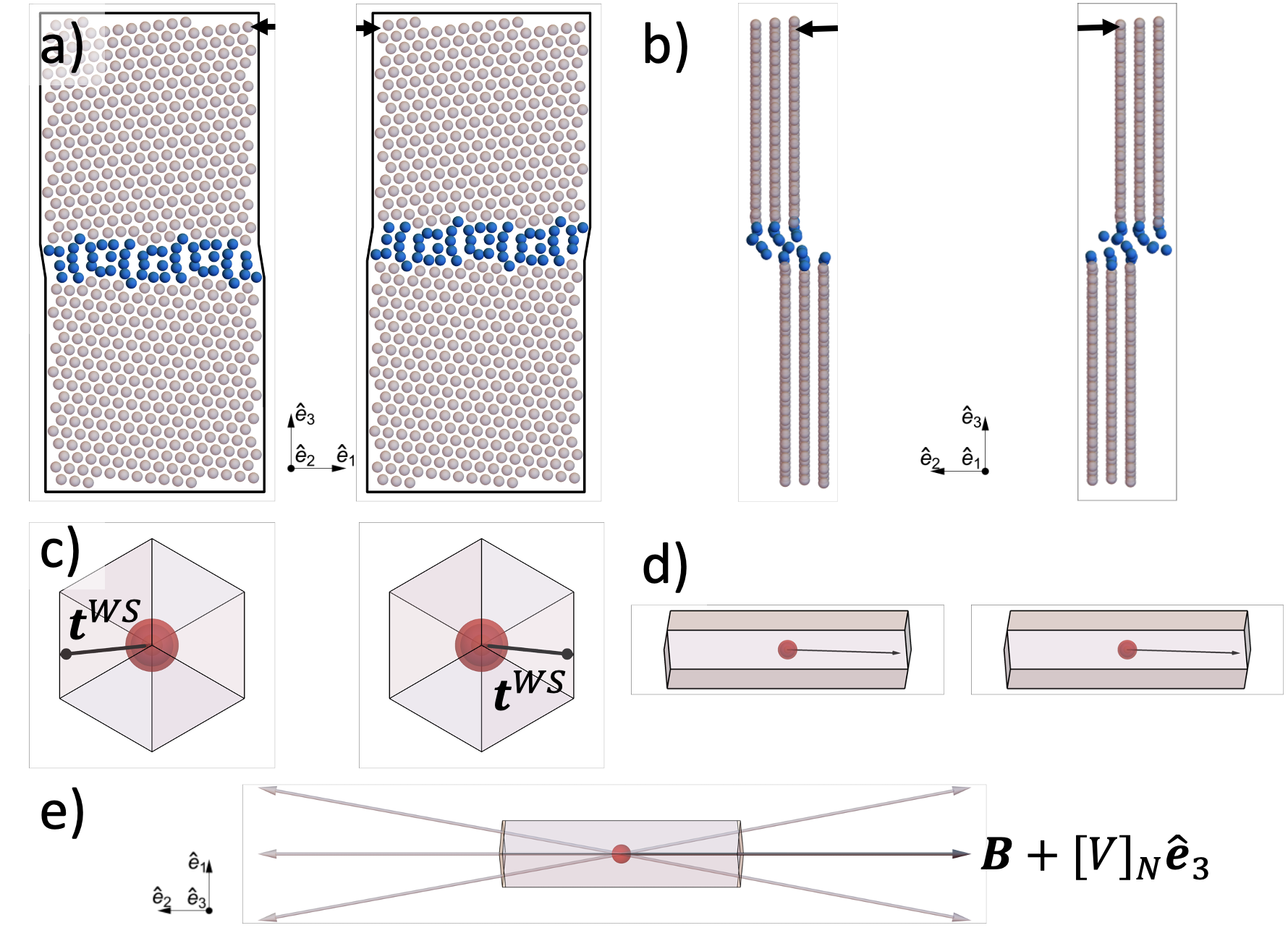}
\caption{Panels a and b: illustration of a $\Sigma 19(178)[11\Bar{1}]$ GB structure that can be produced using different $\gamma$-surface displacements suggesting multiple equivalent definitions of excess shear. Panels c and d: $\bm{t}^{WS}$ is not unique because it lies on the edge of the WS cell of the DSCL and can thus have different orientations while maintaining the same magnitude. Panel e: possible vectors $\bm{t}^{WS}+\bm{d}^{DSC}$ with the third component equal to GB excess volume, with darkest arrow having the smallest magnitude assigned as excess shear, $B$.}
\label{fig:excess_shear}
\end{figure}

In this section, we argue that, unlike the well-defined excess volume, excess shear is not unique and can only be defined up to a DSCL vector. As a result, its proper definition requires a convention. We suggest that the translation vector, $\bm{t}^{WS}$, introduced in this work, can be used to specify such a convention. As an illustration, we consider the $\Sigma 19(178)[11\Bar{1}]$ GB in Cu previously investigated by direct high resolution electron microscopy as well as atomistic simulations \cite{Meiners}. Examples of the GB structure obtained using different $\gamma$-surface translations are illustrated in Fig. \ref{fig:excess_shear}a and \ref{fig:excess_shear}b. We used unwrapped atomic coordinates to show that while the boundary structure is the same for all translations shown, the shape of the bicrystal is different. Based on these bicrystal shapes, we could, in principle, define different GB excess shears, which would lead to an inconsistent thermodynamic treatment. Alternatively, when $\bm{t}^{WS}$ resides inside the WS cell it is unique and can be used to specify unique excess shear values, then all different bicrystal shapes correspond to that unique state plus a disconnection with a DSCL vector.

We selected this boundary to illustrate the most complex case, where even $\bm{t}^{WS}$ is not unique and resides on the surface of the WS cell of the DSCL, allowing it to be located on either face of the WS cell, as shown in Fig. \ref{fig:excess_shear}c. As a result, the $\bm{t}^{WS}$ vector for this boundary structure has two degenerate states. We find that $\bm{t}^{WS} = \pm 0.293$\ \AA$\hat{\bm{e}}_1-0.983$\ \AA$\hat{\bm{e}}_2 - 0.030$\ \AA$\hat{\bm{e}}_3$ and note that the third component does not match the excess volume, $[V]_N = 0.139$\ \AA\ \cite{Meiners}. To find a translation state for which the normal component matches  the sum of the excess volume and the additional displacement due to the GB atoms described by $[n]$, we need to add an additional DSCL vector to Eq. \eqref{eq:t3ws} such that

\begin{subequations}\label{eq:consistent}
    \begin{align}
        t_3^{WS} + d_3^{DSC} &= [V]_N+ n\Omega^b/A,\label{eq:excess-condition}\\
        \bm{d}^{DSC} &= u \bm{a}^{DSC} + v \bm{b}^{DSC} + w \bm{c}^{DSC},\label{eq:define-DSC} 
    \end{align}
\end{subequations}
with $u$, $v$ and $w$ being integers. $\bm{a}^{DSC}$, $\bm{b}^{DSC}$ and $\bm{c}^{DSC}$ are primitive DSCL vectors. As a result of Eq. \eqref{eq:excess-condition} the components of excess shear are given by $B_i = t_i^{WS} + d_i^{DSC}$. Eq. \eqref{eq:consistent} describes an under-determined system of equations. As such, we describe one possible convention to specify a unique excess shear: we define  the integers $u$, $v$ and $w$ such that $B=\sqrt{B_1^2+B_2^2}$ is minimized subject to the constraint imposed by Eq. \eqref{eq:excess-condition}.

For the $\Sigma 19(178)[11\Bar{1}]$ grain boundary in an FCC crystal we can express the primitive DSCL vectors as $\bm{a}^{DSC} = -\frac{a_0\sqrt{38}}{76}\hat{\bm{e}}_1 + \frac{a_0 \sqrt{114}}{76}\hat{\bm{e}}_3$, $\bm{b}^{DSC} = \frac{a_0\sqrt{38}}{76}\hat{\bm{e}}_1 + \frac{a_0 \sqrt{114}}{76}\hat{\bm{e}}_3$, and $\bm{c}^{DSC} = -\frac{a_0\sqrt{38}}{76}\hat{\bm{e}}_1 - \frac{a_0\sqrt{3}}{3}\hat{\bm{e}}_2+ \frac{a_0 \sqrt{114}}{228}\hat{\bm{e}}_3$. The number of GB atoms $[n]=0$ for this particular structure. For a more general case of non-zero $[n]$ a convention has to be specified as $[n]$ can be assigned as positive or negative. In this work, we require $[n]$ to be positive. If we apply Eq. \eqref{eq:consistent} to $\bm{t}^{WS}$, and note that $[V]_N = 0.139$\ \AA, we find the constraint $w = 1-3(u+v)$. By minimizing the magnitude of the excess shear, we find $\bm{B} = -3.070$\ \AA$\hat{\bm{e}}_2$. It is important to note that regardless of the sign of $t_1^{WS}$ we find the same value of excess shear per our convention. Thus, this convention removes the degeneracy seen with $\bm{t}^{WS}$ for this case.

Once a unique excess shear is assigned to a given GB structure the different shapes of the  bicrystals, shown in Fig. \ref{fig:excess_shear}, can be reconciled by assuming that the given translation state is a sum of that unique excess shear and an additional DSCL vector. The thermodynamic equations describing equilibrium coexistence must then be modified accordingly to include this additional Burgers content at the junction due to disconnections. For example, for non-climb cases, the well-known Gibbsian equilibrium condition derived for fluids and requiring the equality of the interfacial free energies of two phases ($\alpha$ and $\beta$) must be modified to

\begin{equation}\label{eq:gibbs-pk}
    \gamma^{\alpha}-\gamma^{\beta} = \sum_{i=1}^2\sigma_{i3}d_i^{DSC}.
\end{equation}
Where both interfacial free energies, $\gamma^{\alpha}$ and $\gamma^{\beta}$, of two different GB phases are calculated according to the specified excess shear convention, which in turn determines the  DSCL Burgers content of the disconnection on the right-hand side of Eq. \eqref{eq:gibbs-pk}. Eq. \eqref{eq:gibbs-pk} shows that the difference in GB free energies is balanced by the Peach-Koehler (PK) force acting on a DSCL disconnection. The PK term on the right-hand side of Eq. \eqref{eq:gibbs-pk} is necessary because, unlike fluid systems, GBs can have multiple different GBPJs between identical GB phases when additional DSCL disconnections are absorbed into the junctions. These reactions only change the Burgers vector of the GBPJs, leaving the two GB phases unaltered. As a result, if thermodynamic equilibrium is established between the two GB phases for a junction without any additional DSCL disconnections such that $\gamma^{\alpha}=\gamma^{\beta}$, other configurations of the same GB phases with additional DSCL disconnections cannot be in equilibrium and will move resulting in a GB phase transition. The driving force (and even its direction) for the GB phase transformation is not equal to the difference in free energies, since $\gamma^{\alpha}=\gamma^{\beta}$, but is given by the work done by the PK force. The general case of Eq. \eqref{eq:gibbs-pk} will be addressed in a separate study.

\section{Defining [\MakeLowercase{\textit{n}}] for Asymmetric Grain Boundaries}

\begin{figure}[ht!]
\centering
\includegraphics[width=\columnwidth]{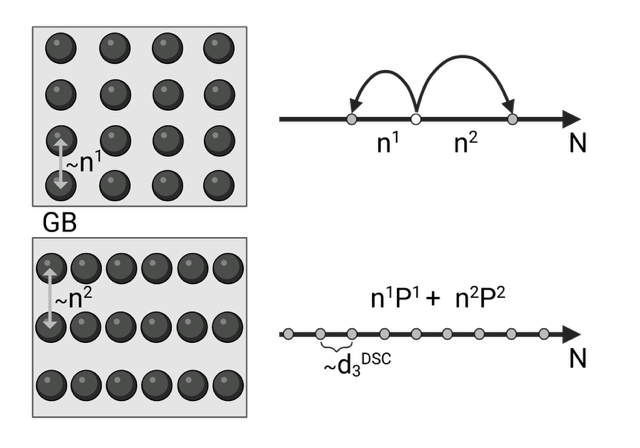}
\caption{ A schematic of two crystals containing different numbers of atoms per plane, $n^1$ and $n^2$, required to construct an asymmetric GB.  All possible bicrystals with different numbers of atoms can be constructed by adding and removing one plane of atoms either $n^1$ or $n^2$ at a time. The set of generated points are equally spaced by the smallest distance given by the greatest common divisor of $n^1$ and $n^2$, which represents the atomic periodicity, $\nu$ of this asymmetric boundary.}
\label{fig:asymmetric}
\end{figure}

While all example GBs considered in this work are symmetric, the calculation of $\bm{t}^{WS}$ can be applied, without modification, to asymmetric GBs as well. As discussed within this work, and in Ref. \cite{FrolovBurgers}, the Burgers vector of a GB phase junction can be expressed in terms of the excess properties $[V]_N$ and $[n]$ of the constituent GB phases. While, the definition of $[V]_N$ does not change between a symmetric and asymmetric GB, we are unaware of a definition for $[n]$ in the case of an asymmetric GB existing in the literature. In this section, we generalize the definition of $[n]$ to general asymmetric grain boundaries and relate it to the component of $\bm{t}^{WS}$ normal to the boundary. Consider a bicrystal composed of two grains such that the number of atoms per plane, $n^1$ and $n^2$, in the upper and lower crystals are different, as schematically illustrated in Fig. \ref{fig:asymmetric}. In a system containing an actual grain boundary structure, the total number of atoms in the region spanned by the crossing vector $\bm{w}$ and grain boundary area $A$ is given by:

\begin{equation}\label{eq:asymmetric_setup}
    N = n^1 P^1 + n^2 P^2 + n,
\end{equation}
Here, $n$ refers to the number of extra atoms due to the GB. The value of $n$ is a periodic function because, if atoms are continually inserted or removed from the GB core, the same structure can be obtained at well-defined intervals of $n$. We aim to determine an expression for the period, $\nu$, and express the unique number of GB atoms $[n]$ for each GB microstate as a fraction of $\nu$.

The reference crystals required to form an asymmetric GB are shown in Figure \ref{fig:asymmetric} and can be considered as stacks of an integer number of planes. Here, the top crystal has $n^1 = 4$ atoms per plane, while the bottom crystal has $n^2 = 6$ atoms per plane. Next, we imagine a random walk in 1D, where at each step, we randomly move left or right by a distance of $n^1$ or $n^2$. Each step is equivalent to adding (positive direction) or removing (negative direction) a plane of atoms containing either $n^1$ or $n^2$ atoms to and from the reference system of two crystals. This random walk produces a set of equally spaced points separated by the greatest common divisor of $n^1$ and $n^2$:

\begin{equation}\label{eq:define-nu}
    \nu = \gcd(n^1,n^2).
\end{equation}
From this treatment, we find that $\nu$ essentially gives us a projection of the DSCL onto the direction normal to the interface. Indeed, the set of points on the grid corresponds to a combination of the two reference crystals, with the number of atoms in the combined system given by:
\begin{equation}\label{eq:diophantine}
    N^0 = n^1 P^1 + n^2 P^2.
\end{equation}
Eq. (\ref{eq:diophantine}) is a linear Diophantine equation. As such, for Eq. (\ref{eq:diophantine}) to have a solution in which $P^1$ and $P^2$ are integers,  $N^0$ must be divisible by the greatest common divisor of $n^1$ and $n^2$ \cite{dudley2012elementary}: $N^0$ must be divisible by $\nu$. If Eq. (\ref{eq:diophantine}) does not have a solution, then $n\neq0$. If Eq. \ref{eq:diophantine} does have a solution, then $n=0$. As a result, $\nu$ represents the atomic periodicity of a given asymmetric GB and defines the number of atoms interval that can give rise to distinct microstates.

We can now assign a unique descriptor for the number of GB atoms of a given GB microstate.  $n$ is the remainder of the division of $N$ by $\nu$, or expressed as a congruence:

\begin{subequations}\label{eq:modulo}
    \begin{align}
        n &\equiv N \ (\mathrm{mod}\ \nu),\\
        [n] &= n/\nu.
    \end{align}
\end{subequations}
For the case of symmetric grain boundaries with $\nu = n^1 = n^2$, we recover the same definition previously introduced in Ref. \cite{Frolov2013}. However, for the more general case of asymmetric boundaries considered here, the atomic interval, $\nu$, can be much smaller than the number of atoms in a single atomic plane of either crystal.

Finally, we relate this new definition of $[n]$ to the normal component, $t_3^{WS}$, and the DSCL. To do so, we note that $n^1$ and $n^2$ are proportional to the separation of interatomic planes in the two crystals: $d^i = n^i \Omega^b /A$. 
Multiplying Eq. \eqref{eq:diophantine} by $\Omega^b /A$ we obtain a projection of the DSCL lattice onto one dimension. As a result, the smallest DSCL component normal to the grain boundary can be expressed as 
\begin{equation}\label{eq:smallest_DSC}
    d_{3}^{min} = \nu \Omega^b/A.
\end{equation}

To relate the number of GB atoms of a GB microstate to its translation vector, we subtract the excess volume contribution from a given value of $\bm{t}$ and then wrap the difference, $\Delta \bm{t} = \bm{t}-[V]_N\hat{\bm{e}}_3$, into the WS cell. We define a wrapping function that takes as an input $\Delta \bm{t}$ and outputs the $\hat{\bm{e}}_3$ component of the wrapped vector, $\Delta t_3^{WS}$. Combining this operation with 
Eq. \eqref{eq:t3ws} and \eqref{eq:smallest_DSC}, we are left with a new expression for the excess number of atoms at a grain boundary:

\begin{equation}\label{eq:nstar}
    [n^*] = \frac{\Delta t_3^{WS}}{d_3^{min}}.
\end{equation}

Eq. \eqref{eq:nstar} provides an alternate definition of the the number of GB atoms. $[n^*]$ is given by the ratio of two vector components: one representing the excess displacement due to GB atoms, $\Delta t_3^{WS}$, and the other associated with the interplanar distance of the DSCL normal to the GB, $d_3^{min}$. The operation of bringing a vector into the WS cell is analogous to the modulo operation in Eq. \eqref{eq:modulo}. The value of $[n^*]$ is equivalent to $[n]$ but it is shifted by $1/2$: $-1/2\leq[n^*]<1/2$ while $0\leq[n]<1$. Negative values of $[n^*]$ correspond to the addition of vacancies to the reference state rather than atoms, which can provide a more descriptive representation for certain GB structures.



\section{Conclusion}

In this work we consider the space of all distinct GB microstates as being   contained within the WS cell of the DSCL of the GB. In this space each GB microstate is represented by a unique GB translation vector, $\bm{t}^{WS}$. This unique vector is obtained by calculating the smallest non-DSCL component of a crossing vector connecting any two lattice sites inside the crystals forming the boundary. $\bm{t}^{WS}$ effectively represents an excess relative translation in three dimensions of the crystals relative to each other due to the presence of the GB. The components of the translation vector are determined by the GB excess properties such as excess shears, excess volume and the number of GB atoms. The components of the vector are equal to those excess properties up to a DSCL vector. This vector can be calculated for a given GB structure without any knowledge regarding how the structure was created from the reference crystals. 

We demonstrate how the $\bm{t}^{WS}$ and the WS space of microstates can be calculated and visualized for a model system represented by a high angle symmetric tilt $\Sigma 5(210)[001]$ GB in FCC Cu. In previous work, we performed rigorous exploration of the structure of this boundary which included both sampling of translations parallel to the boundary as well as the optimization of the number of atoms at the boundary and generated more than a thousand distinct microstates of this boundary. Using these structures, we calculate $\bm{t}^{WS}$ for each of the generated microstates/structures and visualize them inside the WS cell of the DSCL constructed for this bicrystal. This analysis shows that the density of states of $\bm{t}^{WS}$ inside the WS space is highly nonuniform and forms three localized clusters that correspond to three different GB phases of this boundary studied in previous work. Therefore not only does this approach allow us to visualize the microstates of a given boundary, the proposed translation vector should also be considered as a valuable descriptor that can be used by machine learning algorithms to interpret grain boundary structure searches and automatically identify distinct GB phases.

Grain boundary phase junctions are line defects that separate two different GB phases during a GB phase transformation. These defects play the role of interface in the 2D boundary plane and their energy determines the GB transformation barrier as well as the kinetics of such transformations. In this work we show that $\bm{t}^{WS}$ calculated for individual GB phases can be used to predict the Burgers vectors of the GB phase junctions. We also demonstrate that the Burgers vectors connect the centers of the GB phase clusters inside the WS space. The ability to predict the Burgers vectors of GB phase junctions from isolated GB structures without constructing the actual junctions and performing laborious and non-trivial simulations of GB phase transformations allows for automatic high-throughput explorations of these GB properties, including transformation barriers, as a function of the microscopic degrees of freedom.

We quantified GB excess shear and argue that it is defined up to a DSCL
vector, which has implications for the thermodynamic equilibrium conditions. Because the product $\sigma_{i3}B_{i}$ is a term in the thermodynamic expression for $\gamma$, GB free energy itself is not uniquely defined and becomes unique only after a convention is specified. The convention simply describes partitioning between the excess shear term in the expression for $\gamma$ and the remaining DSCL disconnection component, which is assigned to the Burgers vector of the GBPJ. Unlike in fluid systems, the equality of GB free energies is no longer sufficient for thermodynamic equilibrium and the contribution from the DSCL Burgers content of GBPJs has to be taken into account. In this work, we used the CSL configuration as a reference system to define excess shear, which differs from the original study that introduced $\bm{B}$ \cite{PhysRevB.85.224106}. The original study assumed the existence of a unique mapping between a single crystal and a bicrystal with a GB but did not describe how this mapping could be performed.
Even though we argue here that excess shear and even GB free energy itself are not uniquely defined, their absolute values are not important as long as they are computed consistently. It is their difference, as shown in Eq. \eqref{eq:gibbs-pk}, that determines GB phase equilibrium or the driving force for GB phase transformation.

Finally, the introduced metric generalizes the notion of the number of GB atoms $[n]$ beyond symmetric tilt and twist boundaries. We find that the total number of GB atoms is given by the  greatest common divisor of the numbers of atoms in the two bulk planes and it is related to the normal component of the DSCL vector. This finding enables future grand-canonical exploration of the structure of asymmetric boundaries, including mixed tilt-twist boundaries \cite{WAN2024120293}.

\section{Acknowledgements}
This work was performed under the auspices of the U.S. Department of Energy (DOE) by Lawrence Livermore National Laboratory under contract DE-AC52-07NA27344. Prepared by LLNL under Contract DE-AC52-07NA27344. TF was supported by the U.S. DOE, Office of Science under an Office of Fusion Energy Sciences Early Career Award. Computing support for this work came from the Lawrence Livermore National Laboratory Institutional Computing Grand Challenge program. ISW was funded by the LDRD program at Sandia National Laboratories. Sandia National Laboratories is a multi-mission laboratory managed and operated by National Technology \& Engineering Solutions of Sandia, LLC (NTESS), a wholly owned subsidiary of Honeywell International Inc., for the U.S. Department of Energy’s National Nuclear Security Administration (DOE/NNSA) under contract DE-NA0003525.

\bibliography{aapmsamp}
\end{document}